\documentclass[a4paper,fleqn,12pt]{cas-dc}

\usepackage[numbers,sort&compress]{natbib}

\usepackage{microtype}
\usepackage{xurl}
\usepackage[ruled,linesnumbered]{algorithm2e}
\SetKwComment{Comment}{/* }{ */}
\SetKwFor{For}{for (}{) $\lbrace$}{$\rbrace$}

\usepackage{amsmath,amsfonts}

\usepackage{diagbox}

\usepackage{tcolorbox}
\tcbuselibrary{skins}

\newcommand{\best}[1]{\textbf{#1}}
\newcommand{\second}[1]{\underline{#1}}

\newcounter{takeawaycount}
\newenvironment{takeaway}[1][Takeaway]
{%
  \stepcounter{takeawaycount}
  \begin{tcolorbox}[
    enhanced,
    title={#1~\thetakeawaycount},
    fonttitle=\bfseries\sffamily\color{black},
    colbacktitle=blue!10,
    colframe=blue!80!black,
    colback=white,
    arc=5pt,
    boxrule=1pt,
    toptitle=5pt,
    bottomtitle=5pt,
    left=10pt,
    right=10pt,
  ]
}
{%
  \end{tcolorbox}
}

\def\tsc#1{\csdef{#1}{\textsc{\lowercase{#1}}\xspace}}
\tsc{WGM}
\tsc{QE}
\begin{document}
\let\WriteBookmarks\relax
\def\floatpagepagefraction{1}
\def\textpagefraction{.001}

\shorttitle{} 

\newcommand{\sysname}{LiLIS}
\title [mode = title]{{\sysname}: A Lightweight Distributed Learned Index Framework for Spatial Decision Analysis} 

\author{Zhongpu Chen}
\author{Yikai Dong}
\author{Wanjun Hao}

\begin{abstract}
Spatial query and analysis results are often directly applied to decision-making processes such as facility location, proximity resource discovery, accessibility analysis, and risk assessment. Therefore, the efficiency of underlying spatial data access directly impacts the response speed of spatial decision analysis. Existing distributed spatial analysis systems (e.g., Simba, Sedona) already have relatively mature execution frameworks. However, they incur substantial overhead in local index construction and query refinement, especially in read-intensive scenarios. Recent studies have shown that learned indices exhibit considerable retrieval potential in single-machine settings, yet how to integrate them into distributed spatial analysis systems with low modification costs remains unaddressed. In this article, we present {\sysname}, a Lightweight distributed Learned Index prototype for Spatial decision analysis. Without modifying existing execution engines, {\sysname} integrates machine-learned search strategies with spatial-aware partitioning in a distributed framework, and efficiently supports common spatial queries such as point queries, range queries, $k$-nearest neighbor ($k$NN) queries, and spatial joins. Extensive experiments on both real-world and synthetic datasets demonstrate that {\sysname} achieves lower latency across various query types and reduces index construction overhead compared with baseline approaches. These results indicate its potential for improving the responsiveness of read-intensive spatial decision-support workflows.
\end{abstract}


\maketitle

\section{Introduction}

In recent years, learned indices, which utilize machine learning models to predict the data position directly, have paved the way to efficient search with a succinct data structure~\cite{kraska2018case, sun2023learned}. In spatial scenarios, existing studies have proposed various methods based on space-filling curves, sorting mapping, and multi-dimensional distribution learning~\cite{li2020lisa, liu2023efficiently, almamun2024survey, sheng2023wisk, li2024survey}, and these methods can significantly reduce local search overhead in practice. Simply put, a learned index can be seen as a function which is able to map the key (e.g., spatial coordinates in our work) to the position of underlying data directly with a pre-defined error bound. However, most existing learned spatial index studies, such as LISA~\cite{li2020lisa}, RM-SI~\cite{qi2020effectively}, PLATON~\cite{yang2023platon}, and WISK~\cite{sheng2023wisk}, are primarily designed for standalone or centralized settings. Recent work such as SOLAR has started to explore learning-based optimization for distributed spatial joins~\cite{liu2025solar}, but lightweight learned local indexing for general distributed spatial queries remains relatively underexplored.

Although learned spatial indices have demonstrated effectiveness in reducing local search costs~\cite{li2024survey}, incorporating them into distributed spatial analytics pipelines still involves several practical constraints. Existing approaches often rely on relatively complex distribution modeling procedures that introduce additional index construction overhead~\cite{almamun2024survey,liu2025good}, and some model architectures are not naturally compatible with existing partition-based execution mechanisms~\cite{yu2019spatial}. Furthermore, many learned spatial indices are mainly designed for limited query types~\cite{qi2020effectively,zhang2021sprig}, leaving their applicability to more complex analytical tasks such as spatial joins insufficiently investigated~\cite{liu2025solar}. Therefore, an important research question arises: \textbf{Can lightweight learned spatial indices be integrated into distributed spatial analytics frameworks to reduce data-access latency while preserving compatibility with existing execution engines?}

To address this question, this paper presents {\sysname}, a Lightweight distributed Learned Index prototype designed for Spatial decision analysis workflows. The proposed approach preserves the conventional partition-based execution structure while introducing an error-bounded spline interpolation model as a local index within each partition to predict the positions of spatial objects efficiently. This design reduces local scanning overhead and improves query efficiency during distributed spatial analysis. Moreover, by decoupling spatial partitioning strategies from index construction, {\sysname} can flexibly support multiple types of spatial queries, including point queries, range queries, $k$-nearest neighbor ($k$NN) queries, and spatial joins, while maintaining compatibility with existing distributed spatial analytics frameworks. This design makes the proposed approach suitable for latency-sensitive spatial decision-support scenarios that require efficient candidate retrieval without restructuring existing analytics infrastructures.

In summary, our contributions are threefold:

\begin{enumerate}[a)]
    \item We designed {\sysname} to incorporate a low-overhead local learned index into distributed spatial analysis pipelines while preserving the original execution framework.
    \item We implemented efficient distributed spatial algorithms to support common spatial queries, including point queries, range queries, $k$NN, and spatial join based on {\sysname} on Apache Spark. The source code is available at \url{https://github.com/SWUFE-DB-Group/learned-index-spark}.
    \item We conducted extensive experiments on real-world and synthetic datasets under diverse partitioning strategies, data distributions, and workloads, demonstrating the effectiveness of {\sysname} in improving query responsiveness under read-intensive spatial analysis workloads.
\end{enumerate}

The remainder of this article is organized as follows: Section~\ref{sec:related} reviewed the related work. The overview design and key methodology of {\sysname} are described in Section~\ref{sec:design}. We discuss the details of efficient distributed spatial algorithms in Section~\ref{sec:alg}. Section~\ref{sec:exp} presents the results of our experiments, providing insights about the efficiency and scalability of {\sysname}, and Section~\ref{sec:discussion} synthesizes the system-level contributions of {\sysname}, discusses its implications for spatial decision-support workflows, and suggests specific directions for future research.

\section{Related Work}
\label{sec:related}

\subsection{Spatial Decision Support Systems}

SDSSs have long focused on integrating spatial data, analytical models and decision-making processes. Keenan and Jankowski~\cite{keenan2019spatial} systematically reviewed the thirty-year research evolution of SDSSs, and pointed out that the combination of GIS, spatial analysis and decision models has become a vital development trend in this field. Malczewski~\cite{malczewski2006gis} summarized typical tasks such as spatial evaluation, regional screening and scheme ranking from the perspective of GIS and multi-criteria decision analysis, illustrating that numerous decision-making problems require frequent access to spatial objects and their spatial relationships. In recent years, relevant studies have further applied SDSSs to scenarios including emergency assembly site selection~\cite{yildirim2025spatial}, agricultural monitoring and near-real-time early warning~\cite{aleman2025near}, so as to support decision-making procedures such as candidate region identification, resource allocation and risk assessment. These efforts demonstrate that spatial queries are not isolated database operations, but serve as a critical bridge connecting underlying spatial data with high-level decision analysis. Their response efficiency directly determines the usability of interactive analysis and time-sensitive decision-making.

\subsection{Spatial Data Indexing}

A spatial index is a carefully designed structure engineered to manage spatial objects by leveraging their spatial attributes, such as location, geometry, properties, and relationships~\cite{pandey2021good}. Spatial indices are widely studied and commonly implemented in mainstream spatial databases, with tree-based index structures (e.g., KD‑tree~\cite{gutierrez2024ckd}, Quadtree~\cite{park2019hierarchical}, and R-tree~\cite{guttman1984r,beckmann1990r}) being the most commonly used. Tree-based spatial indices are typically designed as hierarchical structures, partitioning space recursively. Consequently, for most queries, they can generally achieve an average-case time complexity of $O(\log{N})$, where $N$ represents the number of spatial objects. In addition to tree‐based index structures, a simple yet effective technique is grid partitioning, which divides the two‐dimensional data space into fixed‐size or adaptively adjusted grid cells~\cite{jin2022gridtuner,li2022grid}.

Conventional spatial indexing schemes mentioned above are typically designed as general-purpose structures without optimizations aligned to particular data distribution patterns or domain-specific characteristics. Consequently, when applied to large-scale, high-dimensional datasets, they frequently suffer from reduced query throughput and incur significant storage and I/O overheads~\cite{ding2020tsunami,almamun2024survey,zhang2023efficient}. Therefore, there is a compelling need for a lightweight yet efficient solution to mitigate these challenges.

\subsection{Learned Indices}
A learned index is an advanced methodology that leverages machine learning models to replace conventional database indexing structures, such as B-trees and hash tables. The core principle of a learned index is to develop a function that maps a search key to the corresponding storage address of a data object, thereby optimizing query performance~\cite{sun2023learned,ferragina2020,liu2024learned}. T. Kraska \emph{et al.}~\cite{kraska2018case} first introduced the concept of learned indexes, and they proposed the Recursive Model Index (RMI), which hierarchically models data distributions through a multi-stage neural network. However, while their foundational work demonstrates that a learned index can deliver up to a threefold reduction in search time and an order-of-magnitude decrease in memory footprint~\cite{kraska2018case}, it remains confined to static, read-only workloads. Subsequently, J. Ding \emph{et al.} introduced ALEX~\cite{ALEX2020}, an updatable learned index that adaptively partitions data to sustain high query and insert performance under dynamic workloads. 

Beyond the one-dimensional learned indices, recent work further extends them into multi-dimensional settings~\cite{liu2025good}. Typically, learned methods first map two-dimensional or multi-dimensional spatial objects into a sortable one-dimensional key space, and then model the mapping relationship between keys and ranks. The Z-order model~\cite{liu2023efficiently} extends RMI to spatial data. This method uses a space-filling curve to order points and learns the Cumulative Distribution Function (CDF) to map each point’s key to its rank. The recursive spatial model index~\cite{qi2020effectively} further develops these principles by employing the rank space–based transformation technique~\cite{qi2018theoretically}. 

\subsection{Distributed Spatial Analytics}
To process large-scale spatial data, a large body of big spatial analytics system have been proposed over the years~\cite{shin2022comparative, XuYWGXG20}. Simba~\cite{xie2016simba} enhances SparkSQL with DataFrame based spatial queries using two-stage resilient distributed dataset (RDD) indexing. SpatialSpark~\cite{alam2022survey} implements node-local R-trees on Spark RDDs to minimize data transfer. GeoMesa~\cite{geomesaHP2025} provides spatio-temporal indexing across multiple database backends with Kafka integration. Among these, Apache Sedona~\cite{sedona} stands as state-of-the-art solution, and it delivers superior performance for large-scale geospatial processing through its innovative distributed indexing and query optimization architecture. Recent studies have further explored learning-based optimization techniques within distributed spatial analytics environments. For example, SOLAR~\cite{liu2025solar} introduces a learning-based framework to improve the efficiency of distributed spatial join processing through workload-aware optimization strategies. These efforts indicate that integrating machine learning techniques into distributed spatial analytics pipelines is becoming an emerging research direction.

Nevertheless, most existing distributed spatial systems still rely on tree-based or grid-based structures at the local indexing layer~\cite{li2024survey}. In cases of skewed data distribution, low query selectivity, or interactive analysis scenarios, the costs of local index construction and retrieval may become critical performance bottlenecks. Accordingly, there is a pressing need to adopt lightweight local learned indexes to replace part of traditional search procedures without modifying the existing execution framework.

\section{{\sysname} Design}
\label{sec:design}

\begin{figure*}[!t]
    \centering
    \includegraphics[width=.8\linewidth]{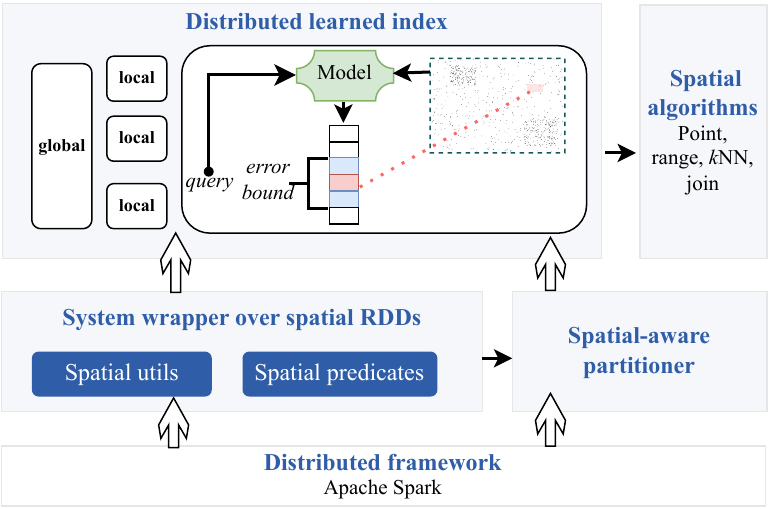}
    \caption{The architecture of {\sysname} on Apache Spark. It introduces spatial RDDs with learned index, and supports a wide range of spatial queries.}
    \label{fig:framework}
\end{figure*}

In this paper, we mainly adopt Apache Spark, one of the most popular distributed computing engines nowadays, as the underlying framework to implement {\sysname} to support spatial points, and the architecture is illustrated in Figure~\ref{fig:framework}. As we can see, it introduces an extra wrapper beyond the RDD so that {\sysname} is able to support (1) spatial functions (e.g., geometry transformation and distance computation); and (2) spatial predicates (e.g., \texttt{contains}, \texttt{intersects} and \texttt{within}). Notably, a learned index component is implemented within partitions of spatial RDDs, which are generated by several spatial-aware partition strategies. By decoupling indexing and partitioning, {\sysname} applies learned models to replace traditional spatial search with a learned alternative while maintaining compatibility with existing components in distributed frameworks, enabling the feasible implementation of a diverse range of spatial query algorithms. The key notations and symbols used throughout this paper are summarized in Table~\ref{tab:notation}. It is worth noting that the overall design of {\sysname} is independent with the specific distributed framework, and the porting to alternatives such as Apache Flink is left as a future work. In what follows, the design of spatial-aware partitioner and learned index are detailed.

\begin{table}[t]
\centering
\caption{Common notations used in this paper}
\label{tab:notation}
\begin{tabular}{lp{0.7\columnwidth}}
\toprule
\textbf{Symbol} & \textbf{Description} \\
\midrule
$R$ & Input spatial RDD \\
$R^*$ & Key-value RDD after partition assignment \\
$o=(x,y,v)$ & Spatial object with coordinates and attribute \\
$G$ & Set of spatial grids from the partitioner \\
$\mathcal{P}$ & Spatial partition \\
$\mathcal{S}$ & Learned spline index model \\
$\hat{p}$ & Predicted position from spline model \\
$\epsilon$ & Prediction error bound \\
$T$ & Radix lookup table \\
$q$ & Query point \\
$r$ & Search radius \\
$k$ & Number of nearest neighbors \\
$d$ & Spatial object density \\
\bottomrule
\end{tabular}
\end{table}

\subsection{Spatial-aware Partitioner}
Typically, data in a distributed system shall be partitioned among machines in a cluster. For the sake of load balance and data locality, data-aware partitioners beyond plain range and hash strategies are designed to boost the performance~\cite{chen2020itiss}. Therefore, {\sysname} adopts the similar method used in Simba~\cite{xie2016simba} to leverage the spatial locality. For the large scale spatial data, we observed that partitioning based on sampling can achieve a great trade-off between efficiency and effectiveness. In this paper, we set sampling rate to 1\% in a uniform way. It is worth noting that partitioner itself can be seen as a \textit{global} index, and {\sysname} provides several built-in grid-based (e.g., fixed and adaptive grid) and tree-based (e.g., R-tree, Quadtree, and KD-tree) spatial-aware partitioning methods to adapt for a wide range of applications. As for tree-based partitioners, {\sysname} only maintains the leaves as the partitions. When there is no ambiguity, the leaves generated from tree-based methods are also called \textit{grids} in this paper. Notably, since R-tree is generally built in a bottom-up way, the sampling-based partitioning may not cover all spatial objects. Therefore, we introduce a novel concept in {\sysname}, dubbed as an \textit{overflow grid}, referring to a special grid for all remaining \textit{overflowed} spatial objects that do not belong to any leaf in R-tree.

Subsequently, each grid is assigned to a unique identifier (an \texttt{int} in this paper) for re-partitioning. Without loss of generality, we assume the partitioner is R-tree based, and the partitioning algorithm is illustrated in Algorithm~\ref{alg:partition}.

\begin{algorithm}[!ht]
\caption{Partitioning algorithm in {\sysname}}\label{alg:partition}
\KwData{Spatial RDD $R$}
$SR \gets$ sampling over $R$\;
Construct grids list $G$ through the specified method over $SR$\;
\Comment*[l]{Outer foreach is a parallel map in Apache Spark}
\ForEach{object $o \in R$} {
    isOverflow $\gets True$\;
    \ForEach{grid $(id, g) \in enumerate(G)$ } {
        \If{$g$ contains $o$} {
            $R^*$.add($id$, $o$)\;
            isOverflow $\gets False$\;
            break\;
        }
    }
    \If{isOverflow} {
     $R^*$.add($|G|$, $o$)\;
    }
}
Re-partition over $R^*$ by keys\;
Map $(id, o)$ to $o$ over $R^*$\;
\end{algorithm}

The outer for-loop (Lines 3-15) is indeed a parallel \texttt{map} operation in Apache Spark, and each spatial object $o$ will be mapped into a tuple $(id, o)$, where $id$ is the identifier of the grid in which $o$ locates. The \textit{enumerate} function borrows from Python (Line 5), emitting a tuple containing an $id$ (starting from 0) and the $grid$ obtained from iterating over $G$. As for \textit{overflowed} objects, its partition $id$ will be assigned to the number of grids (Lines 12-14). Therefore, a spatial RDD $R$ will be transformed into a key-value RDD $R^*$, and a re-partitioning procedure (Line 16) through shuffling can be conducted over $R^*$ by the key (i.e., the $id$ of the partition). Finally, $R^*$ will be mapped back into a spatial RDD with objects only (Line 17). Since the partitioner acts as a global index, the master node must maintain all partition properties within $G$, including their spatial boundaries and identifiers.

\subsection{Distributed Learned Spatial Index}\label{subsec:index}
To keep the compatibility with existing systems~\cite{xie2016simba,yu2019spatial,chen2020itiss}, {\sysname} follows the two-phase filtering solution, and the local index is implemented via a learned model within a given partition. Assume that a spatial object is $(x, y, v)$, where $x$, $y$, and $v$ are its latitude, longitude, and extra attribute (e.g., the textural description), respectively. The two-dimensional spatial points shall be projected into one dimension for sorting, and possible sorting criteria can be either one arbitrary axis or some aggregated value (e.g., Z-order curve and GeoHash). In this paper, we denote the chosen sorting criteria as the \textit{key} of a spatial object, so each spatial point can be represented as a tuple of $(key, x, y, v)$.

\begin{figure}[!t]
    \centering
    \includegraphics[width=\linewidth]{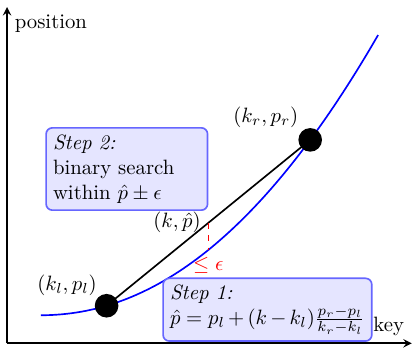}
    \caption{The main idea of spline index is to obtain an estimated position $\hat{p}$ of a given key $k$ based on two adjacent points (Step 1), and then perform a binary search within positions $\hat{p} \pm \epsilon$ in the sorted dataset (Step 2). The overall searching time complexity is constant after retrieving the lower bound $k_l$ and upper bound $k_r$.}
    \label{fig:spline}
\end{figure}

The main idea of the learned spatial index in {\sysname} is to construct a spline model so that it can predict the position of every spatial point with error guarantees~\cite{kipf2020radixspline}. To be specific, given an object $(key, x, y, v)$ whose position in the sorted dataset $\mathcal{D}$ is $p$, the spline model $\mathcal{S}$ can be described as $\mathcal{S}(key) = \hat{p} \pm \epsilon$, where $\epsilon$ is a pre-specified position error bound so that $|\hat{p} - p| \leq \epsilon$. To build the model $\mathcal{S}$ over the sorted dataset $\mathcal{D}$, we need to find a set of representative objects so that the maximum interpolation error is not larger than $\epsilon$, as illustrated in Figure~\ref{fig:spline}. As for the one-pass error-bounded spline algorithms that construct the model, we refer the reader to~\cite{neumann2008smooth}. Given a partition $P$ with $N$ spatial objects, the distributed learned spatial index is conducted by \texttt{mapPartitions} in Apache Spark, without time-consuming data shuffling between machines.

Compared to the commonly-used R-tree index using Sort-Tile-Recursive (STR) algorithm~\cite{leutenegger1997str} whose time complexity is $O(N\log{N} + N\log{f} \times \log_f{N})$, where $f$ is the fanout, the time complexity of the learned index in {\sysname} is $O(N\log{N} + N)$. As for searching, the time complexity of the learned index is $O(\log{M})$, where $M \ll N$ is the size of the set of spline points. As an optimization, the searching time can be further reduced to be constant on average if a radix table is introduced~\cite{kipf2020radixspline}.

\begin{algorithm}[!ht]
\caption{Building a radix table for floating-point keys}\label{alg:radix}
\KwIn{Spline points $S$, number of radix bits $b$}
\KwOut{Radix table $T$}
$T \gets$ new int[$1 \ll (b + 2)$]\;
$min \gets S.first.key$, $max \gets S.last.key$\;
$f \gets (1 \ll b) / (max - min)$\;
$T[0] \gets 0$\;
$prev \gets 0$\;
\ForEach{point $(i, p) \in enumerate(S)$} {
   $curr \gets \text{int} ((p.key - min) \times f)$\;
   \ForEach{$j \in [prev + 1, curr]$}{
      $T[j] \gets i$\;
   }
   $prev \gets curr$\;
}
\ForEach{$prev$}{
    $T[prev + 1] \gets |S| - 1$\;
}
\Return $T$\;
\end{algorithm}

The radix table is proposed to handle the unsigned integer keys (e.g., \texttt{uint32} and \texttt{uint64}) only. To this end, {\sysname} extends this design to support floating-point numbers and strings to fit spatial data. Algorithm~\ref{alg:radix} outlines how to build a radix table for floating-point keys. The size of the radix table is bounded by a user-defined number of bits $b$. Generally, this radix table is used to compress the set of spline points (Lines 6-14) with a scaling factor $f$ (Line 3). Therefore, given a search key $k$, let $k' = \lfloor (k - \mathit{min}) \times f \rfloor$. The lower bound and upper bound positions are $S[T[k']]$ and $S[T[k' + 1]]$, respectively.

\subsection{Discussions on {\sysname} Design}

\subsubsection{Enhancing Rather Than Replacing}
Prior work on learned indices often focuses on replacing traditional indexing structures, which necessitates significant refactoring of system architectures and algorithm implementations. In contrast, {\sysname} adopts a complementary approach by enhancing the existing computation engine without requiring disruptive overhauls, thus reducing adoption barriers for real-world systems.

\subsubsection{Lightweight}
Motivated by the objective of enhancing efficiency, {\sysname} prioritizes a lightweight design. Consequently, constructing a learned index in {\sysname} entails minimal overhead due to its reliance on a limited number of parameters. Furthermore, an additional benefit lies in the ability to adopt a dual-indexing framework, enabling the selection of an appropriate index based on specific requirements. For instance, when either $x$ or $y$ serves as the key, the construction of an R-tree and a learned index can leverage a shared sorting procedure, thereby optimizing the indexing process.

\section{Query Algorithms in {\sysname}}
\label{sec:alg}
In this section, we discuss how to design efficient query algorithms based on the learned index proposed in Section~\ref{sec:design}. 

\subsection{Point Query}
Given a query point $q = (x, y)$ and a dataset $\mathcal{D}$, a point query returns true if $q$ is within $\mathcal{D}$, and returns false otherwise. Obviously, there is at most one partition that contains $q$, and this can be done with RDD's \texttt{filter} method. If the candidate partition exists, then we use the learned index to predict the estimated position, and refine the final result according to Algorithm~\ref{alg:point_query}. As illustrated in Figure~\ref{fig:spline}, we first compute the estimated position (Line 1), and then search the key within the error bound (Line 2). If there is no such key, the algorithm returns false directly (Lines 3-5). Otherwise, we scan in both directions (Lines 6-17).

\begin{algorithm}[!ht]
\caption{Search in point query}\label{alg:point_query}
\KwIn{Candidate partition $\mathcal{P}$, and query point $q$}
\KwOut{True if $q \in \mathcal{P}$, False otherwise}
$\hat{p} \gets \mathcal{P}.learnedSearch(q.key)$\;
$p \gets$ search $q.key$ over $\mathcal{P}[\hat{p} - \epsilon .. \hat{p} + \epsilon]$\;
\If{$p == null$} {
    \Return{False}\;
}
$pos \gets p$\;
\While{$p < \mathcal{P}.size$ and $\mathcal{P}[p].key == q.key$}{
  \If{$q.x == \mathcal{P}[p].x$ and $q.y == \mathcal{P}[p].y$}{
   \Return{True}\;
  }
  $p$++\;
}
$p \gets pos - 1$\;
\While{$p \geq 0$ and $\mathcal{P}[p].key == q.key$}{
  \If{$q.x == \mathcal{P}[p].x$ and $q.y == \mathcal{P}[p].y$}{
   \Return{True}\;
  }
  $p$--\;
}
\Return{False}\;
\end{algorithm}

\subsection{Range Query}
Given a rectangle range $q=(x_l, y_l, x_h, y_h)$, where $(x_l, y_l)$ and $(x_h, y_h)$ are the lower left and upper right points, respectively, a spatial range query returns all points in $\mathcal{D}$ contained in $q$. Similar to the point query, the global filtering can be conducted through a linear scan directly. After obtaining the candidate partitions, we apply \texttt{mapPartitions} to compute the resulting points. The overall procedure is similar to Algorithm~\ref{alg:point_query}. To be specific, we search within the range from lower bound $(x_l, y_l)$ to upper bound $(x_h, y_h)$ in a given partition. In practice, it can be optimized from several aspects. For example, if the partition is enveloped entirely within $q$, then all points should be returned without further checks.

\textbf{Remark 1}: The range query above is indeed a rectangle range query. As for a circle range query with a given center point $p = (x, y)$ and a radius $r$, a practical solution is to construct a minimal bounding rectangle (MBR) for this circle, and perform a regular rectangle range query first. Finally, a subsequent refinement step is necessary to filter out false positives.

\subsection{$k$NN Query}\label{subsec:knn}
Formally, given a query point $q = (x, y)$, a dataset $\mathcal{D}$, and the distance function $d(\cdot, \cdot)$, a $k$NN query returns $\mathcal{R} \subset \mathcal{D}$ so that $|\mathcal{R}| = k$, and for any point $o \in \mathcal{D} \setminus \mathcal{R}$ and any point $p \in \mathcal{R}$, it holds that $d(q, p) \leq d(q, o)$. In {\sysname}, the distributed $k$NN query is implemented with the range query. Firstly, an estimated range is computed according to the following equations~\cite{liu2002efficient}:

\begin{equation}
    r = \sqrt{k / (\pi d)}
\end{equation}
where $d$ is the density of the dataset, given by 

\begin{equation}
    d = N / \texttt{area}
\end{equation}
where $N$ and \texttt{area} are the size and area of $\mathcal{D}$, respectively. 

The first phase of $k$NN search is to conduct a range query with $\hat{q} = (x_q - r, y_q - r, x_q + r, y_q + r)$. If the size of the results is less than $k$, a new searching window is constructed. For $k > 1$, the times of calling the range query are theoretically bounded:

\begin{equation}
    \lceil \frac{\ln{\sqrt{(x_u - x_l)^2 + (y_u - y_l)^2}} - \ln{\sqrt{\frac{k(x_u - x_l)(y_u - y_l)} {\pi N}}}}{\ln{\frac{4k}{\pi(k - 1)}}} \rceil
\end{equation}
where $(x_l, y_l)$ and $(x_u, y_u)$ are the lower bound and upper bound of $\mathcal{D}$, respectively. For the detailed proof, we refer the reader to \cite{liu2002efficient}. In practice, we find that most $k$NN ($k < 10$) can be answered using no more than two range queries.

\subsection{Join Query}
A spatial join takes two datasets ($\mathcal{A}$ and $\mathcal{B}$) and a spatial predicate ($\theta$) as input, and it returns a set of pairs in which the predicate holds, i.e., $\sigma_\theta(\mathcal{A} \times \mathcal{B})$. Due to the diversity of $\theta$, the implementations of join queries exhibit significant variation. In {\sysname}, we currently provide support for the join between polygons $\mathcal{PG}$ and data points $\mathcal{D}$, and the predicate \texttt{contains} predicate checks whether a data point is contained in a polygon. For example, the data points can be shops in a city, and the polygons delineate the commercial zones that are selected on-the-fly. A meaningful join query in this context could aim to identify which shops fall within each commercial zone, enabling an analysis of retail density and distribution patterns. In a practical setting, $|\mathcal{PG}| \ll |\mathcal{D}|$, so polygons can be broadcast to each partition before performing the join. As for the filtering stage, we can compute MBR for each polygon, and then perform a range query whose range is the MBR above to get candidate tuples.

\textbf{Remark 2}: {\sysname} also supports a few other types of joins (e.g., $k$NN-join) out of the box, but how to leverage the learned index to design a unified framework for more types of joins is left as future work.

\section{Experiment}
\label{sec:exp}
In this section, we conduct extensive experiments to demonstrate the superiority of {\sysname} on both real-world and synthetic datasets. All algorithms in this paper are implemented in Java, and all experiments are performed in a cluster with 7 CentOS machines on Tencent Cloud. Each machine is equipped with 8 GB memory, an 80 GB SSD, and an Intel dual-core processor at 3.0 GHz. As mentioned before, the design of {\sysname} is independent of the specific distributed engine, and we present it on Apache Spark 3.0 using Standalone mode, in which one machine is chosen as the master node, and the remaining 6 machines are slave nodes. All configurations in Apache Spark, including garbage collection (GC) and shuffle operations, are maintained at their default settings for the experiments.

\subsection{Setup}
\subsubsection{Datasets}
The datasets used in the experiments are summarized in Table~\ref{tab:dataset}. Both CHI~\cite{chi_data} and NYC~\cite{nyc_data} contain extra information in addition to spatial locations, while SYN~\cite{syn_data} consists of spatial points only, which are generated by Spider. As we can see, although it is possible to load CHI and SYN on a single-node machine, the performance in a distributed setting among different frameworks is a meaningful research question, and these evaluations are sufficient to demonstrate the effectiveness and efficiency of {\sysname}. As for spatial join queries, we constructed 7000 polygons in Chicago, and performed a spatial join with CHI. For other queries, we use NYC as the default dataset. 

\begin{table}[!ht]
\centering
\caption{Spatial Datasets}\label{tab:dataset}
\resizebox{0.98\linewidth}{!}{
\begin{tabular}{l c c l}
\toprule
\textbf{Name} & \textbf{Size} & \textbf{\#items} & \textbf{Description} \\
\midrule
CHI & 1.9 GB & 7M & Crime events in Chicago \\
NYC & 20 GB & 300M & New York taxi rides \\
SYN & 3 GB & 100M & Randomly generated points \\
\bottomrule
\end{tabular}
}
\end{table}

\textbf{Remark 3}: While it is true that the default dataset (i.e., NYC) can be accommodated within the memory of a single contemporary machine, the primary objective of this study is to demonstrate the feasibility and efficacy of {\sysname} within distributed cloud environments. In such settings, cost-effective commodity hardware with limited memory capacity (e.g., 8 GB), as utilized in our experiments, is typically employed, highlighting the system's scalability and performance under resource-constrained conditions.

\subsubsection{Baselines and Implementations}
For evaluation, we compare {\sysname} with Apache Sedona, a representative and widely used distributed spatial analytics framework~\cite{yu2019spatial}. Sedona is shipped with two indices (R-tree and Quadtree) and two partitioners (KD-tree and Quadtree), and we use Sedona-\texttt{\{I\}\{P\}} to denote its various implementation variants, in which \texttt{\{I\}} and \texttt{\{P\}} mean the index type and partitioner, respectively. For example, Sedona-RK adopts the R-tree as the index and the KD-tree as the partitioner. The four variants of Sedona are summarized in Table~\ref{tab:sedona}. Notably, Sedona also supports a no-indexing solution, which we denote as Sedona-N. We denote the vanilla Apache Spark implementation as Spark. As discussed in Section~\ref{sec:design}, {\sysname} can adopt different partitioners. We use {\sysname}-F, {\sysname}-A, {\sysname}-Q, {\sysname}-K, and {\sysname}-R to denote the implementations with fixed-grid, adaptive-grid, quadtree, KD-tree, and R-tree, respectively. In our experiments, we adopt the KD-tree as the default partitioning method for {\sysname}. Consequently, {\sysname}-K serves as the default implementation in our proposed framework.

\begin{table}[!ht]
\centering
\caption{Sedona Variants}\label{tab:sedona}
\resizebox{0.98\linewidth}{!}{
\begin{tabular}{lcc}
\toprule
\diagbox[width=10em]{\textbf{Index}}{\textbf{Partitioner}} & KD-tree & Quadtree \\
\midrule
R-tree   & Sedona-RK & Sedona-RQ \\
Quadtree & Sedona-QK & Sedona-QQ \\
\bottomrule
\end{tabular}
}
\end{table}

Unlike other learned indices that require complex parameters and manual tuning, {\sysname} involves only two hyperparameters: error bound (default: 32) and spline bit count (default: 10). A similar setting is also adopted in \cite{kipf2020radixspline}. This configuration follows the common practice of error-bounded spline indices, aiming to control model complexity rather than pursuing elaborate parameter tuning.

\subsubsection{Workloads}
Unlike the workload in Sedona~\cite{yu2019spatial} which only focuses on different types of spatial queries, in this paper we further consider more subtle factors on range queries~\cite{liu2025good}. To be specific, the selectivity, which is the ratio between the area of a query window and the area of the whole dataset, ranges from $0.00001\%$ (default) to $0.1\%$. In addition, we further adjust skewness for range queries, in which a skewed query (default) follows the distribution of the underlying spatial data, while a uniform query is to construct the input randomly. For $k$NN queries, the default $k$ is 10, ranging from 1 to 1000. To mitigate randomness, all experimental results are averaged over 50 independent runs.

\subsubsection{Research Questions}
In this paper, we mainly focus on the following research questions in terms of {\sysname} through comprehensive experimental evaluations:

\begin{enumerate}[RQ1:]
    \item How does {\sysname} perform on typical distributed spatial queries against baseline competitors under default settings?
    \item What is the effect of different partitioners in {\sysname}?
    \item How robust is {\sysname} across different datasets?
    \item What is the effect of subtle factors on range (e.g., varying selectivities) and $k$NN (e.g., varying $k$) queries in {\sysname}?
    \item What is the cost of building indices in {\sysname}? 
\end{enumerate}

\subsection{RQ1: Overall Performance}
\begin{figure}[!t]
    \centering
    \includegraphics[width=\linewidth]{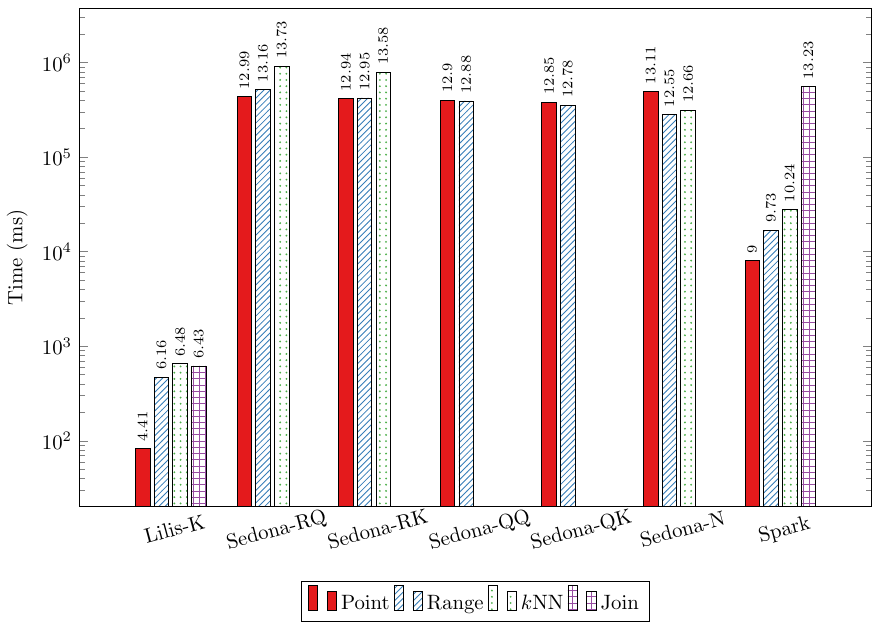}
    \caption{The overall performance under default settings.}
    \label{fig:rq1}
\end{figure}

To answer RQ1, we compare LiLIS-K with the selected baseline implementations, including
Apache Spark and multiple Apache Sedona variants, for range queries,
point queries, and $k$NN queries under the default settings. Because the quadtree index in Sedona \textit{does not support} $k$NN, we only report Sedona-RQ and Sedona-RK for $k$NN queries. For join queries, we use Apache Spark as the baseline, as the evaluated
polygon-point join in LiLIS relies on a customized pipeline that is not
directly comparable with the Sedona variants considered in this study. The evaluation results are illustrated in Figure~\ref{fig:rq1}. As we can see, {\sysname} consistently performs the best compared with other implementations. Since the y-axis in Figure~\ref{fig:rq1} uses a logarithmic scale, the actual performance improvement is greater than the observed values suggest. For example, {\sysname}-K takes about 472 milliseconds for range queries, while Sedona-RQ takes 521282 milliseconds. Interestingly, we can also observe that Sedona variants with indices are not better than Sedona-N and Spark under our settings, and this phenomenon is caused by the underlying data distribution as well as its Filter and Refine model. The authors of Sedona~\cite{yu2019spatial} also verify this finding.

\begin{takeaway}
Under default settings, {\sysname} achieves lower latency across the
tested query workloads, and this advantage is further magnified in read-intensive scenarios.
\end{takeaway}

\begin{figure}[!t]
    \centering
    \includegraphics[width=\linewidth]{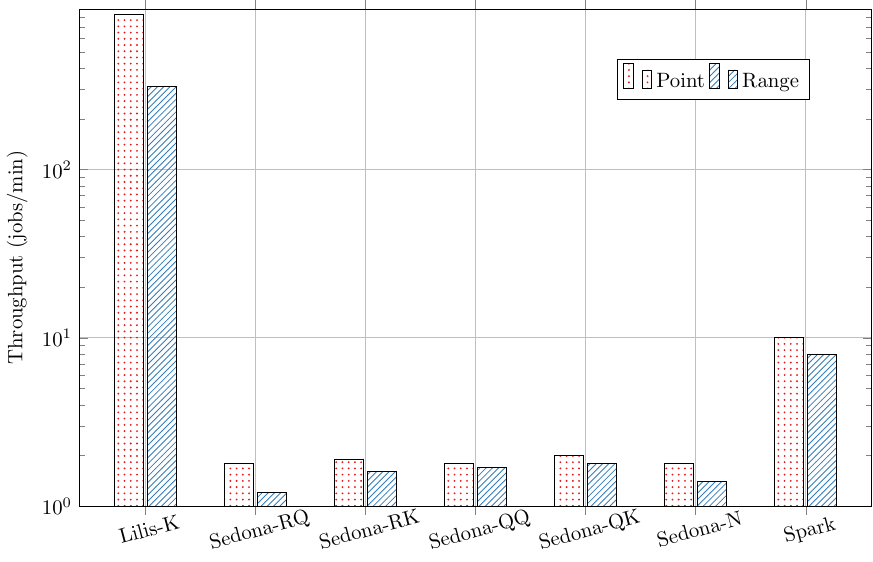}
    \caption{The throughput (jobs per minute) of point and range queries under default settings.}
    \label{fig:throughput}
\end{figure}

As for RQ1, we also conduct the experiments in terms of throughput (\textit{jobs per minute}) of {\sysname} and its competitors on point and range queries. Specifically, we follow similar settings in Simba~\cite{xie2016simba}, and issue 100 queries on the driver program using a thread-pool with 8 threads. The results are shown in Figure~\ref{fig:throughput}. It can be observed that {\sysname}-K consistently achieves higher throughput, but the advantage is not as obvious as that in query time. This is mainly because the throughput is influenced by additional factors such as I/O overhead, CPU cores, and memory bandwidth that are not solely dependent on query execution efficiency.

\subsection{RQ2: Varying Partitioners}

Partitioning strategies directly determine the number, size, and internal data distribution of partitions, thereby greatly affecting the overall performance of {\sysname}. Table~\ref{tab:rq2} presents the average latency of five partitioning schemes across four query types, where the best and second-best results are marked in bold and underlined, respectively.

\begin{table}[!htp]
\centering
\caption{Varying Partitioners (ms)}\label{tab:rq2}
\resizebox{0.98\linewidth}{!}{
\begin{tabular}{lcccc}
\toprule
\diagbox[width=6em]{\textbf{Method}}{\textbf{Query}} 
& Point Query & Range Query & $k$NN Query & Join Query \\
\midrule
\sysname-F & 218.08 & 704.15 & 1107.20 & 636710 \\
\sysname-A & 199.09 & 521.04 & 1767.00 & 1322970 \\
\sysname-Q & 141.30 & \best{340.57} & 773.50 & \second{276557} \\
\sysname-K & \second{82.59} & \second{468.64} & \second{650.20} & \best{228581} \\
\sysname-R & \best{76.68} & 471.55 & \best{618.80} & 21492013 \\
\bottomrule
\end{tabular}
}
\end{table}

The results show that tree-based partitioning outperforms grid-based partitioning overall. This indicates that when partition boundaries better fit the spatial data distribution, the global filtering stage can efficiently narrow down candidate ranges and reduce the overhead of local search. Meanwhile, different query tasks exhibit inconsistent preferences for partitioning strategies. For instance, in join tasks, certain tree-based partitioning schemes benefit single-point location lookup but are less suitable for processing large-scale cross-region candidates. Accordingly, a practical advantage of {\sysname} is that its index design is not tightly coupled with any specific partitioning method, leaving flexible adjustment space for diverse spatial decision analysis tasks.

\begin{takeaway}
{\sysname} is sensitive to various partitioners, and tree-based partitioners generally perform better than grid-based partitioners.
\end{takeaway}

\subsection{RQ3: Varying Datasets}

\begin{figure}[!t]
    \centering
    \includegraphics[width=\linewidth]{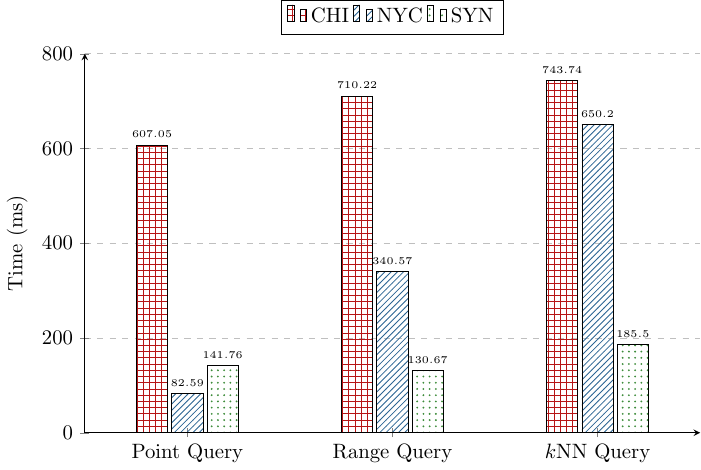}
    \caption{{\sysname} performance while varying datasets.}
    \label{fig:rq3}
\end{figure}

As illustrated in Figure~\ref{fig:rq3}, the experimental results demonstrate significant performance variations across three query types when executed on three distinct datasets. Point queries exhibit substantial performance disparities. Interestingly, despite the NYC dataset being approximately 10 times larger than CHI, it demonstrates significantly better performance for all queries. We can also notice that queries over SYN consistently perform the best, and this is because uniformly generated spatial objects can be best modeled by both spatial partitioners and learned indexes. These findings highlight the importance of considering both dataset size and intrinsic characteristics when optimizing spatial database operations. The results demonstrate that larger datasets do not necessarily lead to proportionally longer query times.

The results in Figure~\ref{fig:rq3} also indicate that $k$NN is more time-consuming than both point and range queries. Therefore, we further take it as a typical case to compare {\sysname} with its competitors across various datasets. In Table~\ref{tab:rq3}, we highlight the best results in bold, demonstrating that {\sysname} consistently outperforms other methods across all datasets.

\begin{table}[!htp]
\centering
\caption{Varying Datasets for $k$NN Queries (ms)}\label{tab:rq3}
\resizebox{0.98\linewidth}{!}{
\begin{tabular}{l c c c}
\toprule
\diagbox[width=10em]{\textbf{Method}}{\textbf{Dataset}} & CHI & NYC & SYN \\
\midrule
\sysname-K & \textbf{743} & \textbf{650} & \textbf{185} \\
Sedona-RK & 7862 & 790993 & 83170 \\
Sedona-N & 6881 & 314243 & 49590 \\
\bottomrule
\end{tabular}
}
\end{table}

\begin{takeaway}
{\sysname} is affected by data characteristics as well as dataset scale, yet our prototype maintains decent adaptability to data with diverse distributions.
\end{takeaway}

\subsection{RQ4: Subtle Factors for Range and $k$NN Queries}
\begin{figure}[!t]
    \centering
    \includegraphics[width=\linewidth]{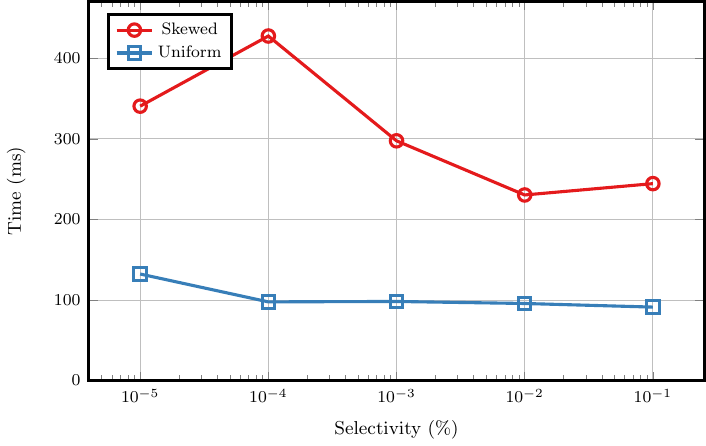}
    \caption{Skewed and uniform range queries under different selectivities.}
    \label{fig:selectivity}
\end{figure}

As described in the experimental setup, both selectivity and skewness are essential for range queries. To explore those subtle factors, we vary selectivity from $0.000001\%$ to $0.1\%$ under both skewed and uniform queries, and the results are reported in Figure~\ref{fig:selectivity}. Uniform queries consistently outperform skewed queries due to their alignment with the underlying data distribution, allowing the learned index to make more accurate predictions. Skewed queries exhibit an inverse relationship between selectivity and execution time. Specifically, higher selectivity often reduces query latency. This occurs because skewed queries with low selectivity values frequently target sparse or irregular data regions, leading to inefficient index traversal. In contrast, as selectivity increases, the query workload becomes more balanced, mitigating the impact of skewness.

\begin{figure}[!t]
    \centering
    \includegraphics[width=\linewidth]{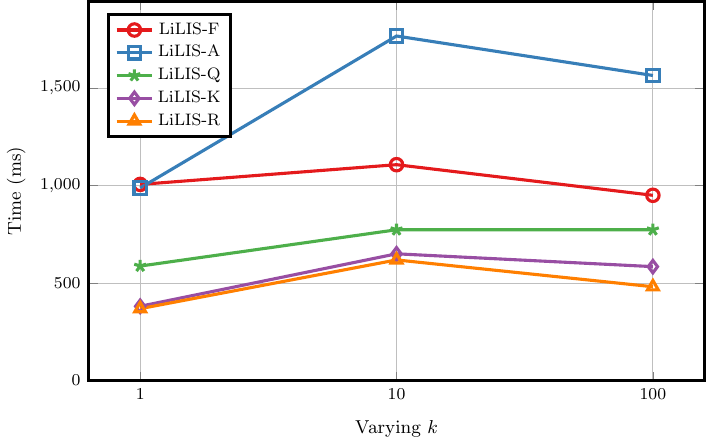}
    \caption{$k$NN queries in {\sysname} when varying $k$.}
    \label{fig:knn}
\end{figure}

Next, we further conduct experiments on all variants of {\sysname} by varying $k$ from 1 to 100 for $k$NN queries. According to the results in Figure~\ref{fig:knn}, we can conclude that $k$NN queries are relatively stable as $k$ varies, and this is consistent with the analysis presented in Section~\ref{subsec:knn}. This stability can be attributed to the query processing mechanism: regardless of how $k$ changes, the number of partitions that need to be accessed remains relatively small, as the query radius determination primarily depends on data distribution rather than the $k$ value itself. Additionally, we can notice that the R-tree partitioner consistently outperforms other partitioning strategies across different $k$ values. This superior performance can be explained by the R-tree's effective spatial indexing capabilities, which enable more efficient pruning of irrelevant partitions during query processing, thereby reducing the overall computational cost.

\begin{takeaway}
Range queries in {\sysname} are sensitive to the skewness, while uniform range queries perform stably across different selectivities; $k$NN queries in {\sysname} are insensitive to common $k$ ($k < 100$).
\end{takeaway}

\subsection{RQ5: Index Cost}

\begin{figure}[!t]
    \centering
    \includegraphics[width=\linewidth]{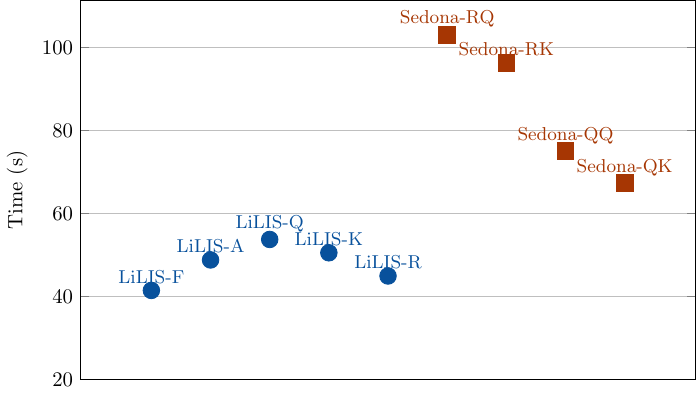}
    \caption{Index building cost.}
    \label{fig:index}
\end{figure}

To demonstrate the superior efficiency of {\sysname} in index construction, we conduct comparative experiments measuring the build time for all index variants in both {\sysname} and Sedona. As illustrated in Figure~\ref{fig:index}, {\sysname} consistently achieves lower index construction overhead compared with existing spatial indices. For instance, building a {\sysname}-K index requires approximately 50 seconds, whereas constructing a Sedona-RQ index takes about 100 seconds, representing a 1.5-2.0$\times$ speedup. Notably, while the performance advantage is significant, the gap in construction time is smaller than the query time improvements we observed above. This observation aligns with our theoretical analysis presented in Section~\ref{subsec:index}. It implies that the practical value of {\sysname} lies primarily in read-intensive analytical scenarios with reusable local indexes, rather than workloads involving frequent data updates or continuous index reconstruction.

\begin{takeaway}
Building indices in {\sysname} is faster than in competitors, though the speed-up is less than its query performance advantage.
\end{takeaway}

\section{Discussion}\label{sec:discussion}

\subsection{System-Level Contributions}

Focusing on the underlying data access challenges in spatial decision analysis, this paper presents {\sysname}, a lightweight distributed learned index prototype that refines the widely adopted two-phase spatial filtering architecture in distributed spatial analytics systems. By combining spatial-aware partition-level pruning with fine-grained local search using learned indexes, {\sysname} supports common spatial analysis operators including point queries, range queries, $k$NN queries, and spatial joins without modifying the underlying distributed execution engine.

Experimental results demonstrate that under read-intensive workloads, {\sysname} achieves lower query latency and reduced index construction overhead compared with existing approaches. These results indicate that lightweight learned local indexes can serve as an effective refinement layer for improving candidate filtering efficiency within distributed spatial query pipelines.

\subsection{Implications for Spatial Decision-Support Workflows}

In many spatial decision-support systems, spatial queries such as range search, nearest-neighbor retrieval, and spatial joins are not standalone operations but intermediate analytical operators repeatedly executed during candidate region screening, accessibility evaluation, and spatial relationship inspection. Therefore, the efficiency of distributed spatial query processing directly affects the responsiveness of iterative spatial analysis workflows.

By reducing partition-level search cost and lowering local index construction overhead, {\sysname} provides an efficient indexing-layer enhancement for read-dominated analytical environments where spatial queries are repeatedly executed under varying analytical constraints. Such characteristics make {\sysname} suitable for integration into spatial decision-support infrastructures that require efficient multi-stage filtering while maintaining compatibility with existing distributed spatial processing frameworks.

Instead of introducing a new execution architecture, {\sysname} improves spatial access efficiency through lightweight refinements to both partition-level pruning and local search stages, making it particularly applicable to exploratory spatial analysis scenarios where users iteratively evaluate alternative spatial conditions during planning and decision-support processes.

\subsection{Limitations and Future Work}

Although the experimental results validate the effectiveness of {\sysname} under read-intensive workloads, several limitations remain to be addressed in future work. First, we will conduct additional experiments on parameter sensitivity, resource overhead and update workloads to fully characterize the applicable scope of our method. Second, we will explore diverse local models beyond linear interpolation to achieve a more robust trade-off among model complexity, error control and decision cost. Third, we will integrate {\sysname} into complex spatial decision-making tasks to further evaluate its effectiveness in terms of end-to-end analysis efficiency and decision support capability.





\bibliographystyle{elsarticle-num-names}

\bibliography{paper}

@misc{sedona,
  author={{Apache Software Foundation}},
  title={{Apache Sedona}},
  year={2025},
  url={https://sedona.apache.org},
  note={Accessed: 2026-05-06}
}

@inproceedings{xie2016simba,
  title={Simba: Efficient in-memory spatial analytics},
  author={Xie, Dong and Li, Feifei and Yao, Bin and Li, Gefei and Zhou, Liang and Guo, Minyi},
  booktitle={ACM SIGMOD International Conference on Management of Data},
  pages={1071--1085},
  year={2016}
}

@article{yu2019spatial,
  title={Spatial data management in apache spark: the geospark perspective and beyond},
  author={Yu, Jia and Zhang, Zongsi and Sarwat, Mohamed},
  journal={GeoInformatica},
  volume={23},
  pages={37--78},
  year={2019},
  publisher={Springer}
}

@inproceedings{beckmann1990r,
  title={The R*-tree: An efficient and robust access method for points and rectangles},
  author={Beckmann, Norbert and Kriegel, Hans-Peter and Schneider, Ralf and Seeger, Bernhard},
  booktitle={ACM SIGMOD International Conference on Management of Data},
  pages={322--331},
  year={1990}
}

@article{li2024survey,
  title={A survey of multi-dimensional indexes: past and future trends},
  author={Li, Mingxin and Wang, Hancheng and Dai, Haipeng and Li, Meng and Chai, Chengliang and Gu, Rong and Chen, Feng and Chen, Zhiyuan and Li, Shuaituan and Liu, Qizhi and others},
  journal={IEEE Transactions on Knowledge and Data Engineering},
  volume={36},
  number={8},
  pages={3635--3655},
  year={2024},
  publisher={IEEE}
}

@inproceedings{kraska2018case,
  title={The case for learned index structures},
  author={Kraska, Tim and Beutel, Alex and Chi, Ed H and Dean, Jeffrey and Polyzotis, Neoklis},
  booktitle={ACM SIGMOD International Conference on Management of Data},
  pages={489--504},
  year={2018}
}

@article{sun2023learned,
  title={Learned index: A comprehensive experimental evaluation},
  author={Sun, Zhaoyan and Zhou, Xuanhe and Li, Guoliang},
  journal={Proceedings of the VLDB Endowment},
  volume={16},
  number={8},
  pages={1992--2004},
  year={2023},
  publisher={VLDB Endowment}
}

@inproceedings{li2020lisa,
  title={LISA: A learned index structure for spatial data},
  author={Li, Pengfei and Lu, Hua and Zheng, Qian and Yang, Long and Pan, Gang},
  booktitle={ACM SIGMOD International Conference on Management of Data},
  pages={2119--2133},
  year={2020}
}

@inproceedings{liu2023efficiently,
  title={Efficiently learning spatial indices},
  author={Liu, Guanli and Qi, Jianzhong and Jensen, Christian S and Bailey, James and Kulik, Lars},
  booktitle={IEEE International Conference on Data Engineering},
  pages={1572--1584},
  year={2023},
  organization={IEEE}
}

@article{sheng2023wisk,
author = {Sheng, Yufan and Cao, Xin and Fang, Yixiang and Zhao, Kaiqi and Qi, Jianzhong and Cong, Gao and Zhang, Wenjie},
title = {WISK: A Workload-aware Learned Index for Spatial Keyword Queries},
year = {2023},
issue_date = {June 2023},
publisher = {Association for Computing Machinery},
address = {New York, NY, USA},
volume = {1},
number = {2},
url = {https://doi.org/10.1145/3589332},
doi = {10.1145/3589332},
journal = {Proc. ACM Manag. Data}
}

@inproceedings{kipf2020radixspline,
  title={RadixSpline: a single-pass learned index},
  author={Kipf, Andreas and Marcus, Ryan and van Renen, Alexander and Stoian, Mihail and Kemper, Alfons and Kraska, Tim and Neumann, Thomas},
  booktitle={International Workshop on Exploiting Artificial Intelligence Techniques for Data Management},
  pages={1--5},
  year={2020}
}

@article{chen2020itiss,
  title={ITISS: an efficient framework for querying big temporal data},
  author={Chen, Zhongpu and Yao, Bin and Wang, Zhi-Jie and Zhang, Wei and Zheng, Kai and Kalnis, Panos and Tang, Feilong},
  journal={GeoInformatica},
  volume={24},
  pages={27--59},
  year={2020},
  publisher={Springer}
}

@article{gutierrez2024ckd,
  title={cKd-tree: A Compact Kd-tree},
  author={Guti{\'e}rrez, Gilberto and Torres-Avil{\'e}s, Rodrigo and Caniup{\'a}n, M{\'o}nica},
  journal={IEEE Access},
  volume={12},
  pages={28666--28676},
  year={2024},
  publisher={IEEE}
}

@inproceedings{ALEX2020,
  author={Ding, Jialin and Minhas, Umar Farooq and Yu, Jia and Wang, Chi and Do, Jaeyoung and Li, Yinan and Zhang, Hantian and Chandramouli, Badrish and Gehrke, Johannes and Kossmann, Donald and Lomet, David and Kraska, Tim},
  title={ALEX: An Updatable Adaptive Learned Index},
  booktitle={ACM SIGMOD International Conference on Management of Data},
  pages={969--984},
  year={2020}
}

@inproceedings{ferragina2020,
  title={Why Are Learned Indexes So Effective?},
  author={Ferragina, Paolo and Lillo, Fabrizio and Vinciguerra, Giorgio},
  booktitle={International Conference on Machine Learning},
  pages={3123--3132},
  year={2020}
}

@article{liu2024learned,
author = {Liu, Qiyu and Han, Siyuan and Qi, Yanlin and Peng, Jingshu and Li, Jin and Lin, Longlong and Chen, Lei},
title = {Why Are Learned Indexes So Effective but Sometimes Ineffective?},
year = {2025},
issue_date = {May 2025},
publisher = {VLDB Endowment},
volume = {18},
number = {9},
issn = {2150-8097},
url = {https://doi.org/10.14778/3746405.3746415},
doi = {10.14778/3746405.3746415},
journal = {Proc. VLDB Endow.},
month = may,
pages = {2886–2898},
numpages = {13}
}

@article{zhang2023efficient,
  title={Efficient Learned Spatial Index With Interpolation Function Based Learned Model},
  author={Zhang, Songnian and Ray, Suprio and Lu, Rongxing and Zheng, Yandong},
  journal={IEEE Transactions on Big Data},
  volume={9},
  number={2},
  pages={733--745},
  year={2023}
}

@article{ding2020tsunami,
author = {Ding, Jialin and Nathan, Vikram and Alizadeh, Mohammad and Kraska, Tim},
title = {Tsunami: a learned multi-dimensional index for correlated data and skewed workloads},
year = {2020},
issue_date = {October 2020},
publisher = {VLDB Endowment},
volume = {14},
number = {2},
issn = {2150-8097},
url = {https://doi.org/10.14778/3425879.3425880},
doi = {10.14778/3425879.3425880},
journal = {Proc. VLDB Endow.},
month = oct,
pages = {74–86},
numpages = {13}
}

@article{liu2025good,
  title={How good are multi-dimensional learned indexes? An experimental survey},
  author={Liu, Qiyu and Li, Maocheng and Zeng, Yuxiang and Shen, Yanyan and Chen, Lei},
  journal={The VLDB Journal},
  volume={34},
  number={2},
  pages={1--29},
  year={2025},
  publisher={Springer}
}

@article{almamun2024survey,
author = {Al-Mamun, Abdullah and Wu, Hao and He, Qiyang and Wang, Jianguo and Aref, Walid G.},
title = {A Survey of Learned Indexes for the Multi-dimensional Space},
year = {2025},
issue_date = {March 2026},
publisher = {Association for Computing Machinery},
address = {New York, NY, USA},
volume = {58},
number = {4},
issn = {0360-0300},
url = {https://doi.org/10.1145/3768575},
doi = {10.1145/3768575},
journal = {ACM Comput. Surv.},
}

@inproceedings{zhang2021sprig,
  title={SPRIG: A learned spatial index for range and kNN queries},
  author={Zhang, Songnian and Ray, Suprio and Lu, Rongxing and Zheng, Yandong},
  booktitle={International Symposium on Spatial and Temporal Databases},
  pages={96--105},
  year={2021}
}

@inproceedings{leutenegger1997str,
  title={STR: A simple and efficient algorithm for R-tree packing},
  author={Leutenegger, Scott T and Lopez, Mario A and Edgington, Jeffrey},
  booktitle={International Conference on Data Engineering},
  pages={497--506},
  year={1997},
  organization={IEEE}
}

@inproceedings{neumann2008smooth,
  title={Smooth interpolating histograms with error guarantees},
  author={Neumann, Thomas and Michel, Sebastian},
  booktitle={British National Conference on Databases},
  pages={126--138},
  year={2008},
  organization={Springer}
}

@inproceedings{liu2002efficient,
  title={Efficient k nearest neighbor queries on remote spatial databases using range estimation},
  author={Liu, Danzhou and Lim, Ee-Peng and Ng, Wee-Keong},
  booktitle={International Conference on Scientific and Statistical Database Management},
  pages={121--130},
  year={2002},
  organization={IEEE}
}

@article{park2019hierarchical,
  title={A hierarchical binary quadtree index for spatial queries},
  author={Park, Kwangjin},
  journal={Wireless Networks},
  volume={25},
  pages={1913--1929},
  year={2019},
  publisher={Springer}
}

@inproceedings{guttman1984r,
  title={R-trees: A dynamic index structure for spatial searching},
  author={Guttman, Antonin},
  booktitle={ACM SIGMOD International Conference on Management of Data},
  pages={47--57},
  year={1984}
}

@article{yang2023platon,
author = {Yang, Jingyi and Cong, Gao},
title = {PLATON: Top-down R-tree Packing with Learned Partition Policy},
year = {2023},
issue_date = {December 2023},
publisher = {Association for Computing Machinery},
address = {New York, NY, USA},
volume = {1},
number = {4},
url = {https://doi.org/10.1145/3626742},
doi = {10.1145/3626742},
journal = {Proc. ACM Manag. Data},
month = dec,
articleno = {253},
numpages = {26}
}

@article{qi2020effectively,
  title={Effectively learning spatial indices},
  author={Qi, Jianzhong and Liu, Guanli and Jensen, Christian S and Kulik, Lars},
  journal={Proceedings of the VLDB Endowment},
  volume={13},
  number={12},
  pages={2341--2354},
  year={2020},
  publisher={VLDB Endowment}
}

@article{qi2018theoretically,
  title={Theoretically optimal and empirically efficient r-trees with strong parallelizability},
  author={Qi, Jianzhong and Tao, Yufei and Chang, Yanchuan and Zhang, Rui},
  journal={Proceedings of the VLDB Endowment},
  volume={11},
  number={5},
  pages={621--634},
  year={2018},
  publisher={VLDB Endowment}
}

@article{pandey2021good,
  title={How good are modern spatial libraries?},
  author={Pandey, Varun and van Renen, Alexander and Kipf, Andreas and Kemper, Alfons},
  journal={Data Science and Engineering},
  volume={6},
  number={2},
  pages={192--208},
  year={2021},
  publisher={Springer}
}

@article{shin2022comparative,
  title={A comparative experimental study of distributed storage engines for big spatial data processing using GeoSpark},
  author={Shin, Hansub and Lee, Kisung and Kwon, Hyuk-Yoon},
  journal={The Journal of supercomputing},
  volume={78},
  number={2},
  pages={2556--2579},
  year={2022},
  publisher={Springer}
}

@article{alam2022survey,
  title={A survey on spatio-temporal data analytics systems},
  author={Alam, Md Mahbub and Torgo, Luis and Bifet, Albert},
  journal={ACM Computing Surveys},
  volume={54},
  number={10s},
  pages={1--38},
  year={2022},
  publisher={ACM New York, NY}
}

@misc{geomesaHP2025,
  author       = {{LocationTech GeoMesa Project}},
  title        = {GeoMesa: Store, index, query, and transform spatio-temporal data at scale},
year = {2025},
  url = {https://www.geomesa.org/},
  note         = {Accessed: 2026-05-06},
}

@inproceedings{jin2022gridtuner,
  title={Gridtuner: Reinvestigate grid size selection for spatiotemporal prediction models},
  author={Jin, Jiabao and Cheng, Peng and Chen, Lei and Lin, Xuemin and Zhang, Wenjie},
  booktitle={IEEE International Conference on Data Engineering},
  pages={1193--1205},
  year={2022},
  organization={IEEE}
}

@article{li2022grid,
  title={Grid Adaptive Bucketing Algorithm Based on Differential Privacy},
  author={Li, Xiangjun and Zhao, Xuewen and Zhang, Huijuan and Han, Jideng},
  journal={Mobile Information Systems},
  volume={2022},
  number={1},
  pages={6988976},
  year={2022},
  publisher={Wiley Online Library}
}

@article{XuYWGXG20,
  author       = {Yang Xu and
                  Bin Yao and
                  Zhi{-}Jie Wang and
                  Xiaofeng Gao and
                  Jiong Xie and
                  Minyi Guo},
  title        = {Skia: Scalable and Efficient In-Memory Analytics for Big Spatial-Textual Data},
  journal      = {IEEE Transactions on  Knowledge and Data Engineering},
  volume       = {32},
  number       = {12},
  pages        = {2467--2480},
  year         = {2020} 
}

@article{keenan2019spatial,
title = {Spatial Decision Support Systems: Three decades on},
journal = {Decision Support Systems},
volume = {116},
pages = {64-76},
year = {2019},
issn = {0167-9236},
doi = {10.1016/j.dss.2018.10.010},
url = {https://www.sciencedirect.com/science/article/pii/S0167923618301672},
author = {Peter Bernard Keenan and Piotr Jankowski}
}

@article{malczewski2006gis,
author = {Jacek Malczewski},
title = {GIS‐based multicriteria decision analysis: a survey of the literature},
journal = {International Journal of Geographical Information Science},
volume = {20},
number = {7},
pages = {703--726},
year = {2006},
publisher = {Taylor \& Francis},
doi = {10.1080/13658810600661508},
URL = {    
        https://doi.org/10.1080/13658810600661508
}
}

@Article{yildirim2025spatial,
AUTHOR = {Yildirim, Ridvan Ertugrul and Sisman, Aziz},
TITLE = {Spatial Decision Support for Determining Suitable Emergency Assembly Places Using GIS and MCDM Techniques},
JOURNAL = {Sustainability},
VOLUME = {17},
YEAR = {2025},
NUMBER = {5},
ARTICLE-NUMBER = {2144},
URL = {https://www.mdpi.com/2071-1050/17/5/2144},
ISSN = {2071-1050},
DOI = {10.3390/su17052144}
}

@article{aleman2025near,
title = {A near real-time spatial decision support system for improving sugarcane monitoring through a satellite mapping web browser},
journal = {Smart Agricultural Technology},
volume = {12},
pages = {101084},
year = {2025},
issn = {2772-3755},
doi = {10.1016/j.atech.2025.101084},
url = {https://www.sciencedirect.com/science/article/pii/S277237552500317X},
author = {Bryan Alemán-Montes and Pere Serra and Alaitz Zabala and Joan Masó and Xavier Pons},
}

@misc{chi_data,
  author = {{[dataset] City of Chicago}},
  title = {Chicago Open Data Portal},
  year = {2023},
  howpublished = {\url{https://data.cityofchicago.org/}},
  note={Accessed: 2026-05-06}
}

@misc{nyc_data,
  author = {{[dataset] NYC Taxi and Limousine Commission}},
  title = {TLC Trip Record Data},
  year = {2023},
  howpublished = {\url{https://www.nyc.gov/site/tlc/about/data.page}},
  note={Accessed: 2026-05-06}
}

@misc{syn_data,
  author = {{University of California, Riverside}},
  title = {Synthetic Spatial Dataset Generator},
  year = {2016},
  howpublished = {\url{https://spider.cs.ucr.edu/}},
  note={Accessed: 2026-05-06}
}

@misc{liu2025solar,
  title={SOLAR: Scalable Distributed Spatial Joins through Learning-based Optimization},
  author={Yongyi Liu and Ahmed Mahmood and Amr Magdy and Minyao Zhu},
  year={2025},
  doi={10.48550/arXiv.2504.01292},
  note={arXiv preprint arXiv:2504.01292},
  url={https://arxiv.org/abs/2504.01292}
}



\end{document}